\documentclass[12pt,a4paper]{article}
\usepackage[dvips]{graphics}	
	
\newcommand{\ybox}[2]   {	
 \begin{center}	
 \resizebox{!}{#1\textheight}	
{\includegraphics{#2.eps}}	
 \end{center}           }

\begin{document}	
\thispagestyle{empty}	

\begin{center}	
	
{\LARGE \bf A new approach to inferring the mass composition of cosmic rays at
energies above 10$^{18}$ eV} 
\end{center}	
%
\begin{center}	
{\bf  M. Ave$^1$, J.A.~Hinton$^{1,2}$, R.A.~V\'azquez$^3$, \\	
A.A.~Watson$^1$, and E.~Zas$^3$}\\	
$^1$ {\it Department of Physics and Astronomy\\	
 University of Leeds, Leeds LS2 9JT ,UK \\}	
$^2$ {\it Enrico Fermi Institute, University of Chicago, \\ 	
5640 Ellis av., Chicago IL 60637, U.S.\\}	
$^3$ {\it Departamento de F\'\i sica de Part\'\i culas,\\	
Universidad de Santiago, 15706 Santiago de Compostela, Spain\\}	
\end{center}

\begin{abstract}	

We describe a new approach to establishing the mass composition at high energies. Based
on measuring both the vertical and inclined shower rates, it has the potential
to distinguish heavy nuclei from light nuclei.  We apply the method to Haverah
Park data above $10^{18}~$eV to show that, under the assumption that the Quark
Gluon String Jet Model correctly describes the high energy interactions, the
inclined shower measurements favour a light composition at energies above
$10^{19}$~eV.  The same conclusion is obtained using a variety of assumptions
about the cosmic ray spectrum. To the extent that precise spectral
measurements will be possible by forthcoming experiments such as the Auger
observatories, the method will further constrain data on composition of
the ultra high energy cosmic rays.

\end{abstract}	

{\small {\it keywords}: mass composition : ultra high energy cosmic rays}

\section{Introduction}	
	
Efforts to understand the origin of cosmic rays at any energy are greatly	
hampered by our lack of knowledge of the mass distribution in the incoming	
cosmic ray beam. The determination of the arrival direction of cosmic rays	
does not depend on knowledge of the mass of the primary and, using the	
fluorescence technique, the primary energy is obtainable with only a small	
systematic uncertainty because of the unknown mass.  Even with the traditional	
ground arrays it has been possible to devise ways of deducing the primary	
energy that are reasonably independent of model and mass uncertainties, at	
least at the 30 \% level.  However use of the data on the energy spectrum and	
arrival direction distribution to decide between various origin models does	
require knowledge of the primary mass distribution.  While there is a common	
assumption that protons dominate at the highest energies, hard experimental	
evidence is lacking, as it is at energies above 10$^{18}$ eV.	
	
Some models of cosmic ray origin lead to the conclusion that photons	
dominate the cosmic ray beam.  This hypothesis can be more readily	
tested than can competing hypotheses that advocate protons or iron	
nuclei as the majority primaries.  This is because the shower	
development of a photon primary is very different from that of any	
hadronic primary.  The difference is larger than is expected between	
the showers created by proton and iron primaries.  For example, Halzen	
{\it et al.} \cite{vazquez} have shown that a photon is extremely unlikely	
to produce a shower in the atmosphere that has a development profile	
that looks like the famous event of $3 \times 10^{20}$ eV observed by the	
Fly's Eye Group \cite{320EeV}.  A study of horizontal showers	
recorded at Haverah Park has shown that at 10$^{19}$ eV there can be	
no more than 40 \% of the primaries that are photons \cite{PRL}, while	
a search, using the traditional idea of photon showers being deficient	
in muons, has given a similar limit at the same energy	
\cite{Teshima01}.  These analyses all assume that photons $>$ 10$^{19}$	
eV have not acquired hadron-like properties, as has sometimes been	
speculated.	
	
There is no reliable evidence on the hadronic mass composition above 10$^{19}$
eV.  Some years ago an analysis of the Fly's Eye data \cite{gaisser} pointed
to a change from an iron-dominated composition at $3 \times 10^{17}$ eV to a
proton-dominated composition near 10$^{19}$ eV.  This conclusion was drawn
from a study of the variation of depth of maximum with energy (the elongation
rate) and from an analysis of the spread in depth of maximum at a given
energy.  The elongation rate behaviour was in agreement with an earlier,
model-independent, analysis from Haverah Park \cite{watsonwalker} and with a
related study using Cherenkov light by the Yakutsk group \cite{Yakutsk}.  The
conclusions on mass composition from the study of the spread in depth of
maximum were confirmed by \cite{Wibig} but none of these analyses extended to
energies above 10$^{19}$ eV.
\begin{figure}	
\ybox{0.4}{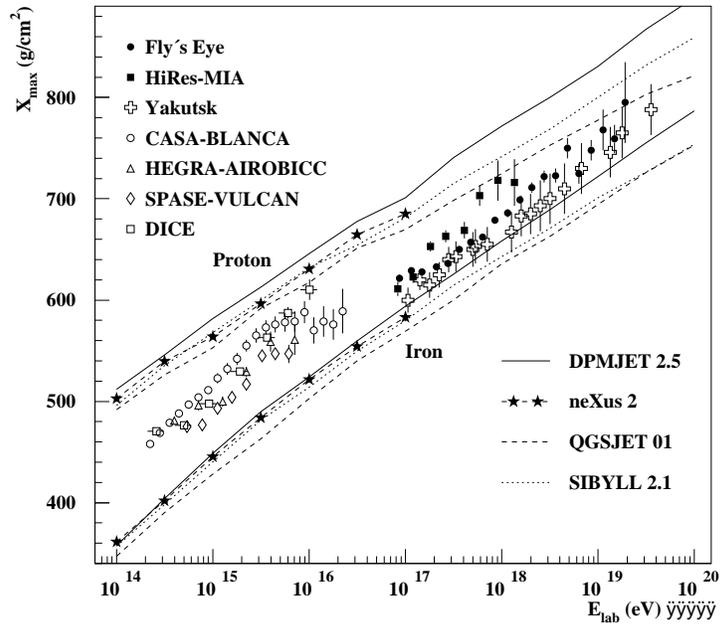}   
\caption{Compilation of data on depth of maximum as a function of energy from	
different experiments compared with predictions for different models. This
figure was supplied by D. Heck and J. Knapp and appear in \cite{heck}}	
\label{xmax}	
\end{figure}	

A different conclusion has been reached by the AGASA group based on the
variation of the muon content of showers with energy \cite{MuAgasa}.  Their
analysis favours a composition that remains 'mixed' over the 10$^{18}$ to
10$^{19}$ eV decade.  Furthermore there is a difference between the
conclusions reached from the data on depth of maximum from the Fly's Eye
experiment \cite{gaisser} and those reached from the HiRes prototype, operated
with the MIA detector, in the range 10$^{17}$ to 10$^{18}$~eV
\cite{HiResMiaDepth}. The more recent data are used to infer protons as early
as 10$^{18}$ eV, if the QGSJET model\cite{QGSJET} is correct. There is a good
evidence from Haverah Park data \cite{compICRC} that the QGSJET model is
satisfactory, at least to 10$^{18}$ eV. However the HiRes/MIA conclusion is,
in turn, in contradiction with a new analysis of Haverah Park data
\cite{compICRC} that suggests that protons make up only 34 \% of the flux,
independent of the energy, between $3 \times 10^{17}$ and $2 \times 10^{18}$
eV, with iron nuclei making up the remainder. 
	
The addition of more data in recent years and a better understanding of what
might be the optimum shower model has not helped to clarify the position:
rather the reverse.  In figure \ref{xmax} we show a compilation of data on
depth of maximum, $X_{\rm m}$, as a function of energy \cite{heck}.  The sensitivity of a
conclusion to the choice of model is evident.  The matter is thus far from
being resolved and there is clearly scope for alternative approaches.

\section{Outline of a new approach}	
	
It seems unlikely that a single method of deriving the mass composition will
be sufficient to resolve this important issue.  In this paper we argue for a
new approach that is based on the study of inclined showers, and which can be
added to the armoury that can be used.  The method is derived from the work
that led to our estimate \cite{PRL} of a limit to the photon flux at 10$^{19}$
eV, an estimate that has now been confirmed independently \cite{Teshima01}, as
mentioned above.  Our approach stems from the detailed understanding of the
propagation of inclined showers that has been developed \cite{model} and from
its application to the prediction of the rate of triggering of the Haverah
Park array \cite{rate}.  Inclined showers induced by very high energy cosmic
rays produce particle densities at ground level which are dominantly due to
muons, in contrast to vertical showers that have an important component due to
photons and electrons stemming from $\pi^0$ decay. Measurements of the cosmic
ray rate at different zenith gives a handle on the relative number of muons
in a shower, which is dependent on composition.
	
To find the photon limit we assumed that the energy spectrum was known, and
calibrated in a mass-independent way, through the work with the Fly's Eye
detector.  Using this known 'vertical' spectrum, a prediction was made of the
number of showers expected with energies above 10$^{19}$ eV for photon, proton
and iron primaries.  We found strong evidence to reject a photon intensity at
the level predicted by several super-heavy relic models
\cite{bkv},\cite{birkel},\cite{rubin}.  Recent reassessment of the energy
spectrum in the appropriate range \cite{Wolfendale}, and the differences
reported between the AGASA group and the HiRes group
\cite{Sakaki},\cite{HiRes}, make it premature to be too firmly in favour of
protons or iron above 10$^{19}$ eV, but our conclusion about photons remains
robust.  An extensive description of our method used to derive the photon
limit has been given recently \cite{hplp}.  In this paper we show how the
technique can be used to infer the mass at energies between 10$^{18}$ and
10$^{19}$ eV.  The conclusions, as to the mass variation with energy that we
can presently draw, is constrained by uncertainties in the form of the
spectrum in this energy range, and by the assumed hadronic interaction model,
rather than by statistical limitations.  We pressent conclusions for a number
of spectra, using the QGSJET model \cite{QGSJET} as a reference for hadronic
interactions, in the expectation that the spectrum uncertainties will soon be
resolved.
	
\section{The analysis of inclined showers}	
	
The analysis of showers with zenith angles $>$ 60$^0$ is not straightforward,
even with an array like Haverah Park in which the detectors were deep-water
Cherenkov detectors.  These detectors present a significant area to inclined
showers by comparison with thin scintillators.  There are two major
difficulties.  Firstly the showers lose their near-circular symmetry very
rapidly above 60$^0$ \cite{Andrews} and, secondly, the finite size of the
array makes it difficult to locate the core.  During the operational phase of
the Haverah Park array (1967 -1987) a decision was made to ignore events $>$
60$^0$ for these reasons.  More recently, with the advent of the analytical
techniques of Ave {\it et al.}  \cite{model} and with the availability of much
more computing power, it has proved possible to get accurate directions for
the events, to estimate their energies (subject to assumptions about shower
models) and to compute the flux.  In this section we justify these claims.
	
The methods of finding the directions and of calculating the energy are
described in detail in \cite{hplp} for events above 10$^{19}$ eV.  Here we
have applied the same techniques, namely an iterative process in which arrival
directions are fitted to the time distributions, followed by shower energy and
core position fits to the detector densities. The timing of the signals use
the information on core position to account for curvature corrections.
Following the same steps as described in \cite{hplp}, we have performed three
quality cuts to eliminate uncertainties in the reconstruction procedure: (i)
the distance from the central triggering detector to the core position in the
shower plane is required to be below $r_{\rm max}=500$~m, (ii) the $\chi^2$
probability for the energy and direction fits must be $>$ 1 \%, (iii) the
downward error in the energy determination is required to be less than a
factor of 2.  The chosen value of $r_{max}$ guarantees that the core position
is always close to the triggering detectors in the array. Similar results,
but with reduced statistics, was obtained with $r_{\rm max}=300$~m. This last
cut is the only one that differs from those used in \cite{hplp}: it ensures
that the smaller showers, that are the concern of this paper, are near to
sufficient detectors to allow reliable density fits.  After making the cuts
described above we found 385 events with $E_0 > 10^{18}$~eV and 2 events with
$E_0 > 10^{19}$~eV.

\begin{figure}	
\ybox{0.3}{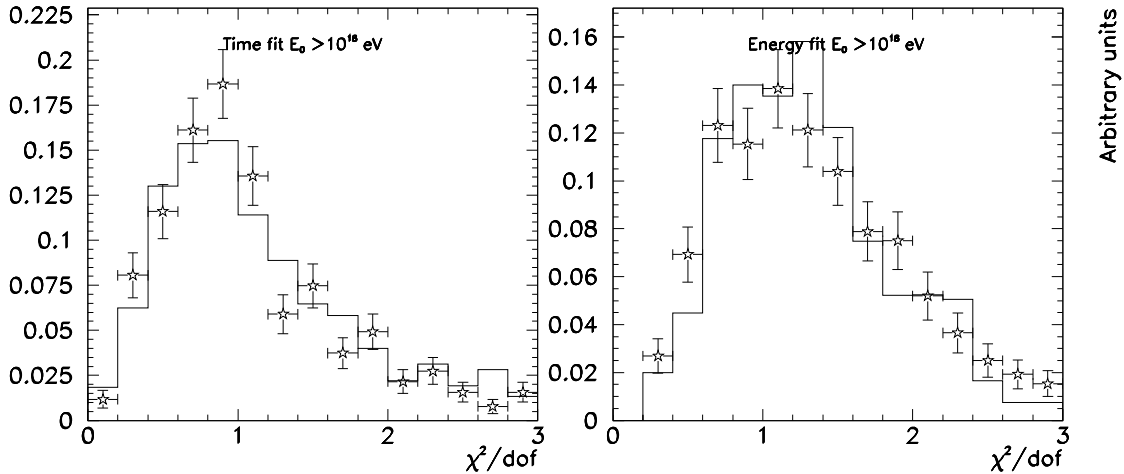}  
\caption{$\chi^2$ distributions from the energy and direction reconstruction	
of data (stars) and artificial events (histogram), assuming proton 	
composition and the parameterizations of the spectrum given 	
in \cite{WatsonNagano}. 	
}	
\label{chis}	
\end{figure}	
Again mimicking the procedure described in Ref.~\cite{hplp}, we have generated
a sample of artificial events to which we have applied the same reconstruction
algorithms as used for the data.  In figure \ref{chis} we show the
normalised $\chi^2$ distributions for the timing fit to the 385 events above
10$^{18}$ eV and for the energy estimates made, assuming proton primaries. The
match between the real data and artificial data generated using the spectrum
description given in \cite{WatsonNagano} (from here on NW), and assuming
proton primaries, is reassuring.  We argue that these plots show
that we have developed a good understanding of the techniques needed to
analyse the directions and energies of very inclined showers.

\begin{figure}	
\ybox{0.3}{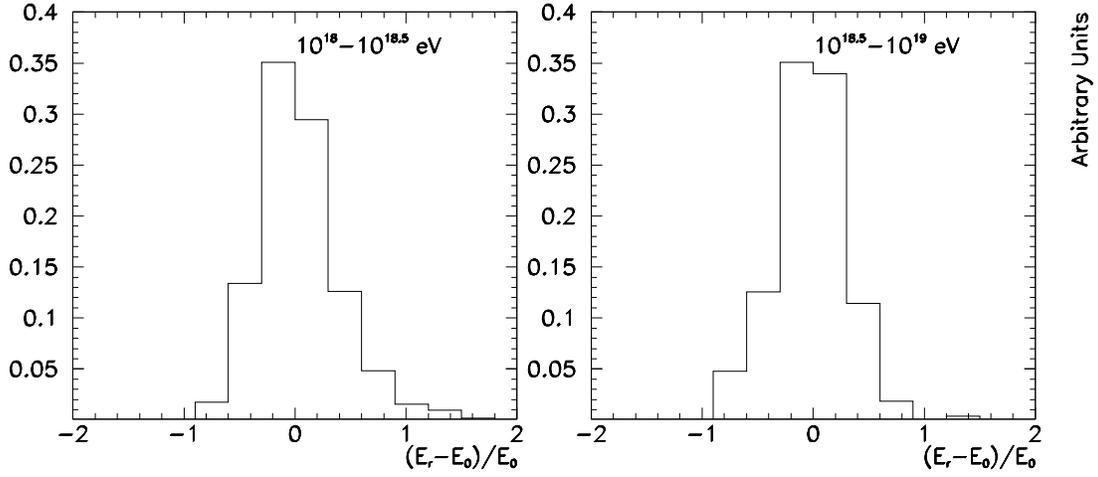}    
\caption{Energy resolution integrated for all zeniths in different energy 
bins. A uniform energy distribution is assumed for each graph.}	
\label{energycomp}	
\end{figure}	

The energy resolution for events above 10$^{18}$ eV is shown in figure
\ref{energycomp} for two energy bands. The energy error for the artificial
events is the difference between the reconstructed energy $E_r$ and that of
the simulated shower $E_0$.  The rms spread in the relative errors for the
artificial events are 0.43 and 0.40: we note the tail towards higher
energies in the lower energy band.

\begin{figure}	
\ybox{0.3}{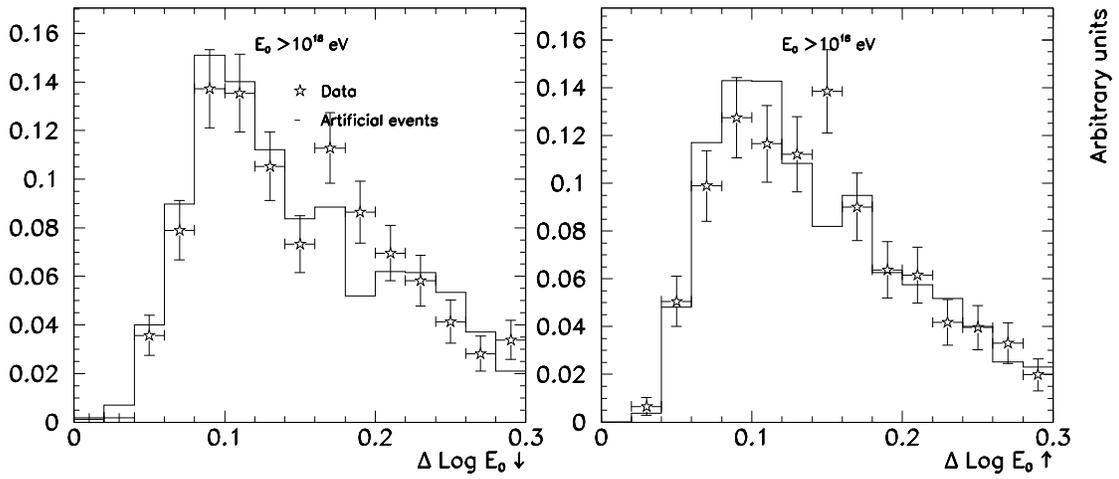}  
\caption{Downward and upward error distribution in the reconstructed 
energy from the density fits to the data (stars) and to the artificial 
events (histogram).} 
\label{errors}
\end{figure}

The energy error in the data is obtained by combining in quadrature 
the error obtained from the maximum likelihood analysis and the 
error associated with the uncertainty in the zenith angle reconstruction. 
In figure \ref{errors} we show the upward and downward error distributions 
for the real data as compared with the artificial events, again generated 
with the NW spectrum and proton primaries. Again the agreement is very satisfactory.

\begin{figure}	
\ybox{0.3}{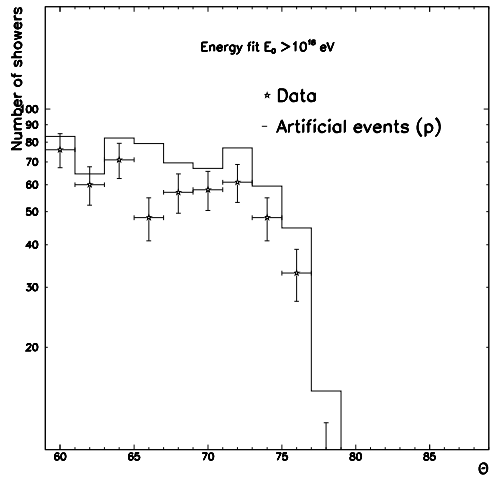}  
\caption{Zenith angle distribution for data (stars) and artificial events	
(histogram). No normalization has been made. Statistical error bars are also shown.}	
\label{zendist}	
\end{figure}

In figure \ref{zendist} the zenith angle distribution for events above
10$^{18}$ eV is compared to real data using the NW (proton) spectrum to make
the prediction.  This comparison is similar to that made in our earlier work
\cite{PRL} but we have now developed methods to estimate the energy, most
importantly including zero density measurements in the energy reconstruction
fits \cite{hplp}.  Here we see a difference between the predictions and the
data.  Note that the data are not normalised. The fact that the predicted rate
is above the artificial rate indicates that we are using (i) a spectrum with
fluxes that are too high, or (ii) the wrong mass composition, or (iii) an
incorrect hadronic interaction model. Of course a combination of all these
uncertainties can also be involved.

\section{Rate dependence on primary mass and spectrum assumptions} 	

We have studied how different assumptions about the shape of primary 
spectrum affect the expected  rates of inclined showers for both pure proton and 
pure iron samples. 
\begin{figure}	
\ybox{0.6}{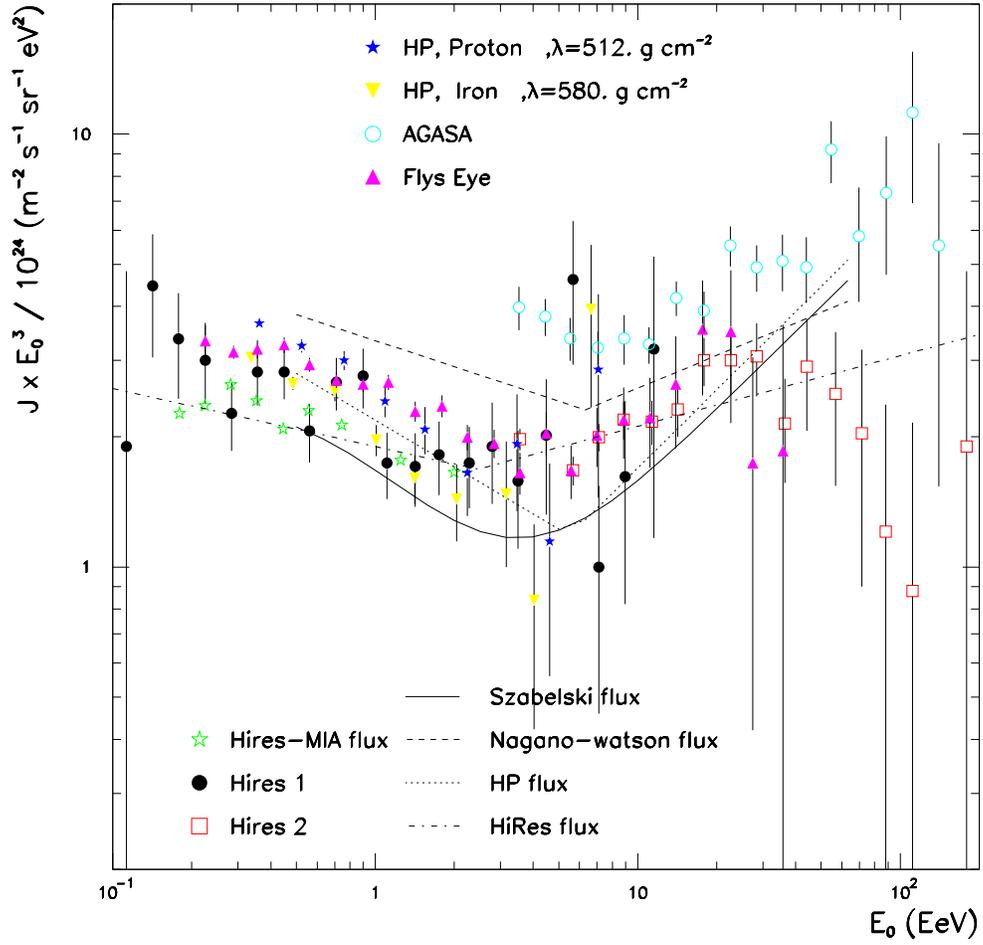}   
\caption{Compilation of cosmic ray data for energies exceeding 
$10^{17}$~eV including, recently reanalysed Haverah Park data assuming 
proton and iron primaries, Stereo Fly's Eye data, Monocular HiRes data 
from both Eyes (HiRes1 and HiRes2), recently reanalysed AGASA 
data, including events up to $60^{\circ}$, and hybrid HiRes MIA data.  
The measurements are compared to spectrum parameterizations given 
by different authors and used in the simulation of artificial events 
as described in the text.}	
\label{fluxes}	
\end{figure}	
In figure \ref{fluxes} we show a compilation of cosmic ray data for energies
above 10$^{17}$ eV together with various parameterizations of the energy
spectrum.  The AGASA \cite{Sakaki}, Monocular HiRes (HiRes 1 and HiRes 2
\cite{HiRes} ), HiRes-MIA \cite{HiResMia}, Haverah Park \cite{specICRC} and
Fly'sEye Stereo \cite{FlysEyeStereo} spectral measurements are as reported by
the different groups. For the purposes of this work we have taken four recent
parameterizations of the spectra represented by the lines in
Fig. \ref{fluxes}: the spectra attributed to Nagano and Watson
\cite{WatsonNagano}, the one obtained by Szabelski {\it et
al.}~\cite{Wolfendale} (both of these are syntheses of various data sets), the
parameterization given by HiRes experiment \cite{HiRes}, and a
parameterization based on a recent analysis of the Haverah Park data
\cite{specICRC}. We have not used the recently reported AGASA spectrum
\cite{Sakaki} which starts above 3 x 10$^{18}$ eV.
\begin{figure}	
\ybox{0.3}{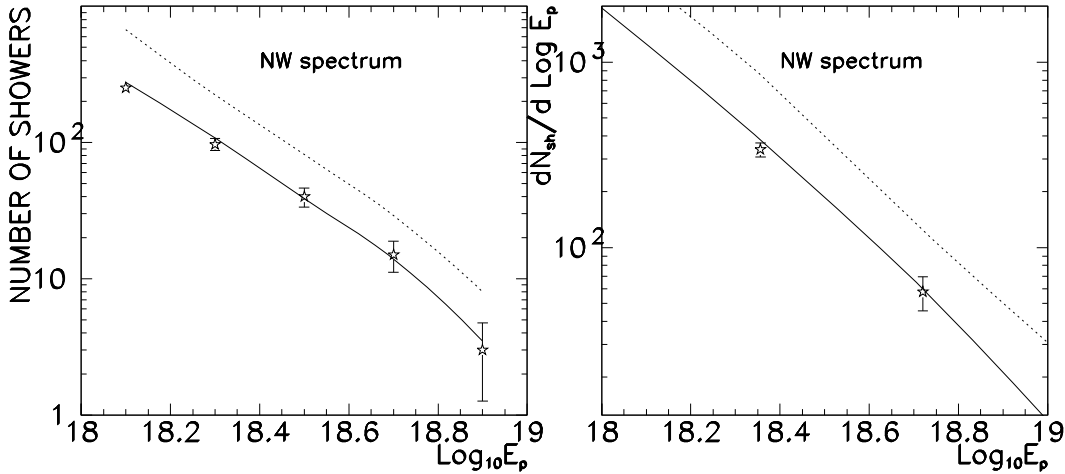}    
\caption{Integral (left panel) and differential (right panel) 	
number of inclined events as a function of energy for the Haverah Park 	
data set (stars) compared to the predictions for iron (dotted line), 	
protons (continuous). The parameterization 	
of the spectrum given in \cite{WatsonNagano} is assumed.}	
\label{nw}	
\end{figure}	
\begin{figure}	
\ybox{0.3}{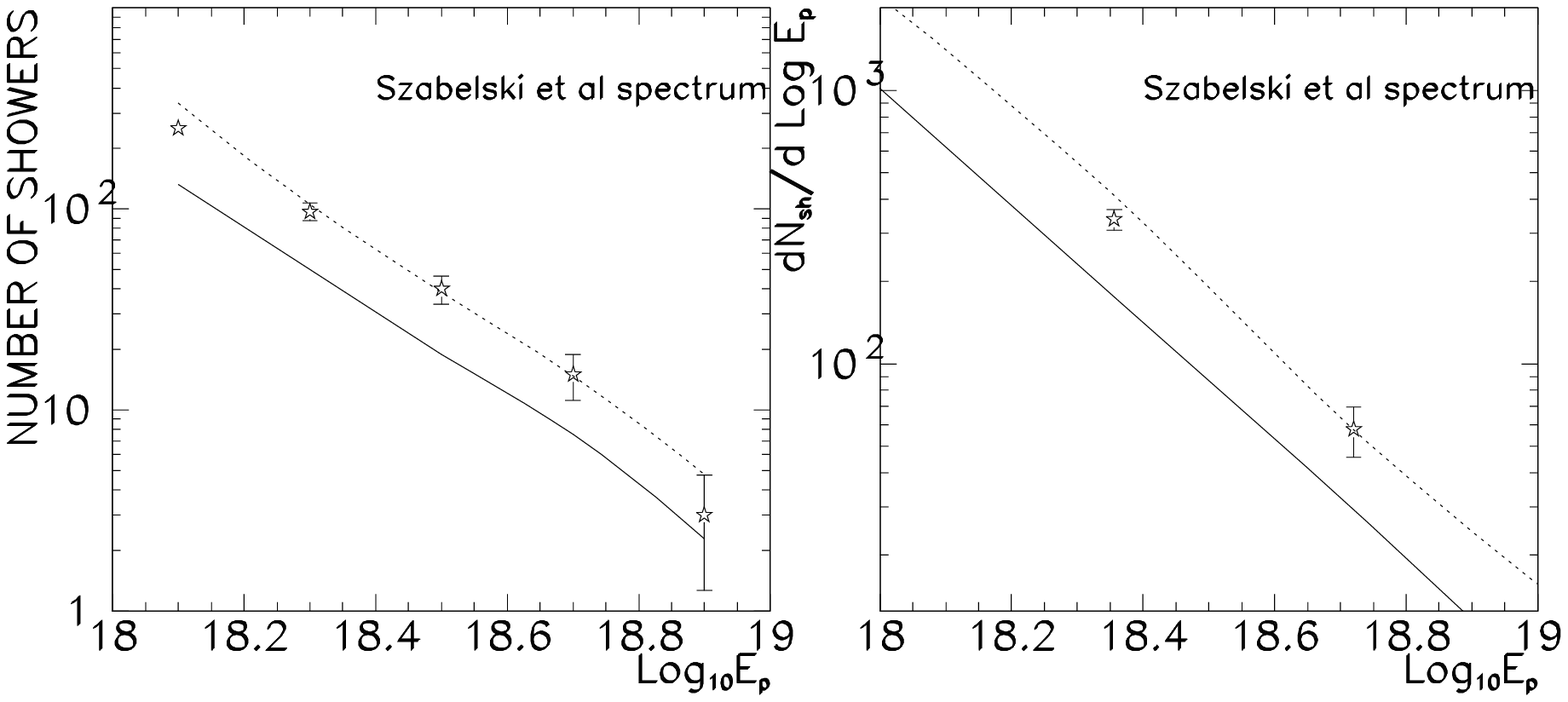}    
\caption{Same as Fig. \ref{nw} but using the parameterization of the	
spectrum given in \cite{Wolfendale}.}	
\label{sza}	
\end{figure}	
\begin{figure}	
\ybox{0.3}{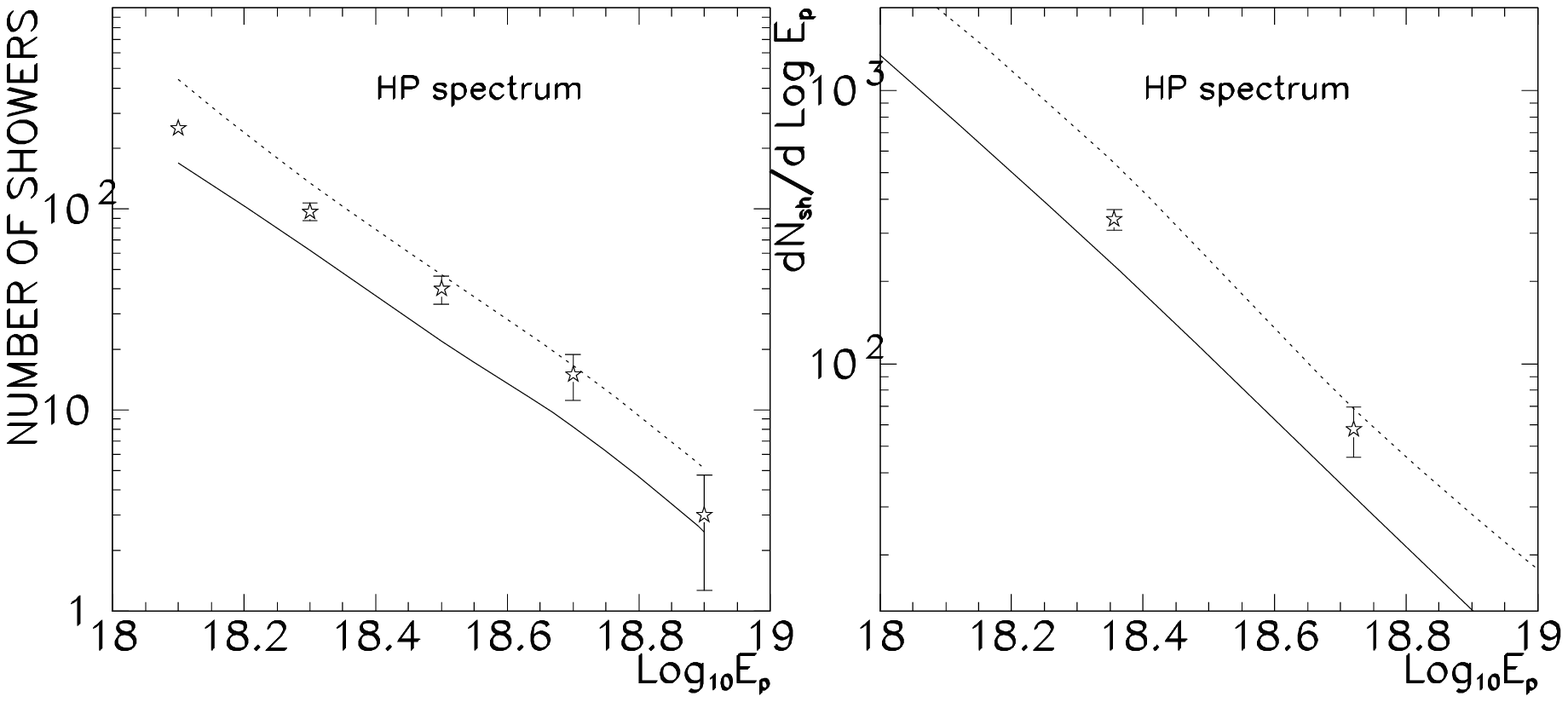}    
\caption{Same as Fig. \ref{nw} but using the parameterization of the	
spectrum given in \cite{specICRC}.}	
\label{hp}	
\end{figure}	
\begin{figure}	
\ybox{0.3}{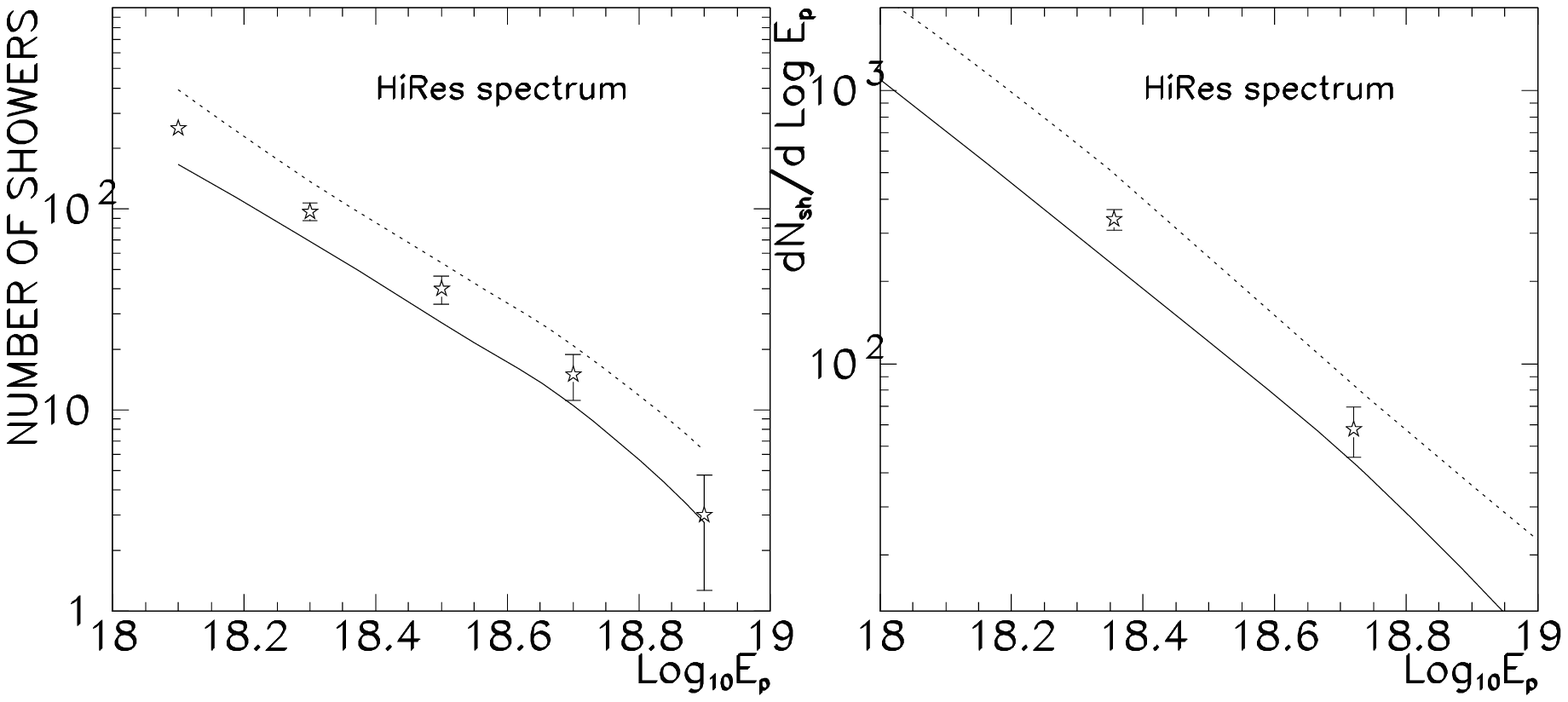}    
\caption{Same as Fig. \ref{nw} but using the parameterization of the	
spectrum given in \cite{HiRes}.}	
\label{hi}	
\end{figure}	
In figures \ref{nw},\ref{sza},\ref{hp},\ref{hi} we show the sensitivity of the
inferred variation of mass for reasons explained later with energy to the
different input spectra.  While from the Szabelski {\it et al.}  input we find
a mass spectrum that is iron dominated over the whole energy range
(10$^{18}$-10$^{19}$ eV), with the NW spectrum one would argue for a proton
dominated composition.  However if we take the recently revised Haverah Park
spectrum \cite{specICRC} we find that the mean mass lies between proton and
iron near 10$^{18}$ eV and possibly gets heavier as we move to energies close
to 10$^{19}$~eV.  We note that the mean mass at the differential energy point
at $2 \times 10^{18}$ eV is $<\ln A> = 1.4 \pm 0.4$. A similar conclusion can
be drawn with the HiRes spectrum as an input. In the energy decade below 2 x
10$^{18}$ eV we have recently argued that $<\ln A> = 2.8 \pm 0.4$, assuming a
bi-modal mass composition of proton and iron \cite{compICRC}.

\begin{figure}
\ybox{0.6}{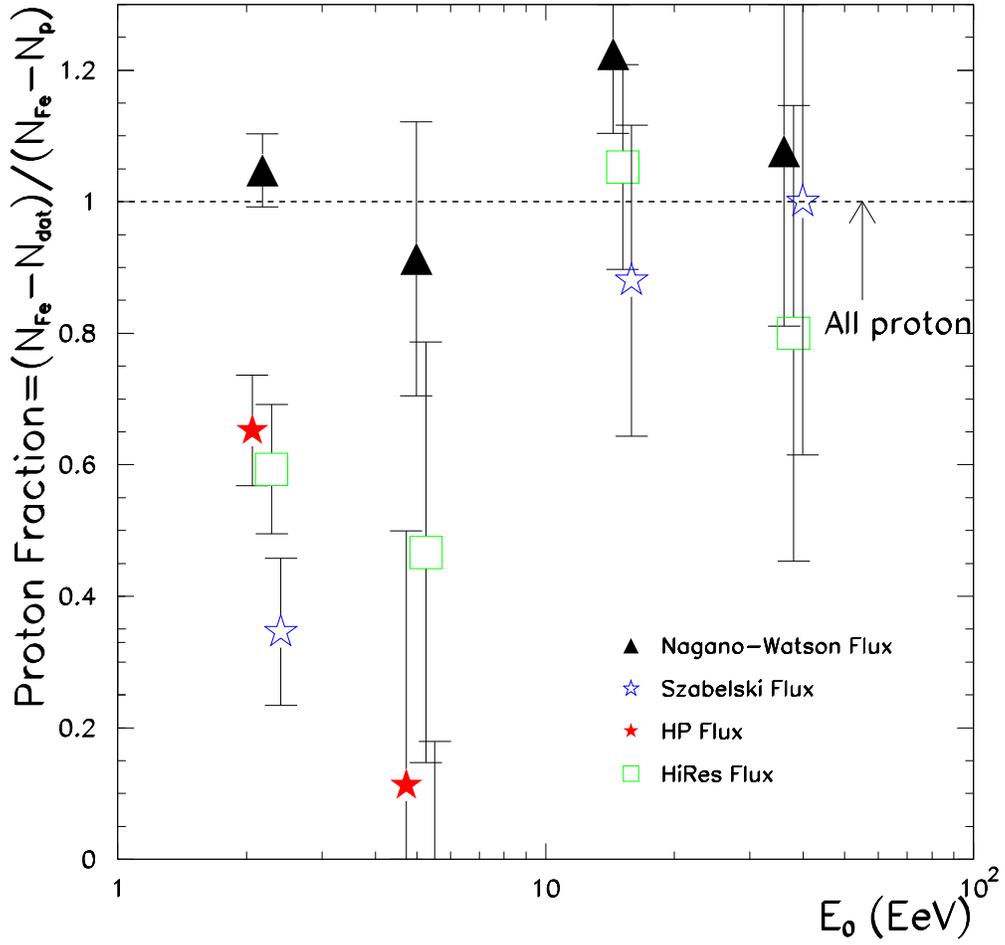}    
\caption{Ratio of proton to total as obtained from a two component mixture 
of protons and irons and assuming four different energy spectra 
parametrizations as shown in figure~\ref{fluxes} and described in the text.}
\label{ratio}
\end{figure}

Comparing the measured data to the corresponding predictions assuming an iron
only spectrum and a proton only spectrum, it is straight forward to obtain the
fraction of protons in a dual mass compositon model.  In figure \ref{ratio}
the variation with energy of the predicted fraction of protons in the cosmic
ray beam is plotted for the four spectra. In this figure, above 10$^{19}$ eV,
we have included events with less stringent quality cuts ($r_{max}=2$~km), as
used in \cite{hplp}.  For sufficiently high energies, the number of detectors
with signals increases substantially for any surface array and this allows the
relaxation of the condition, while retaining reasonably constrained fits.
There are large uncertainties, but a trend towards a light composition at
energies above the ankle is apparent with all spectra used.  We note also that
if the true flux is like that in the recent analysis presented by AGASA
\cite{Sakaki}, the inclined shower data would be inconsistent with even a
dominantly iron composition because the predicted rate would exceed the
observed rate. A heavier mass primary composition would be required.

\section{Conclusion}	
We believe that we have demonstrated the potential of a new method for
extracting the mass spectrum above 10$^{18}$ eV by studying cosmic rays at
high zenith angles.  It is based on the difference in  abundance of muons with
respect to photons and electrons for different primaries.  We are not making
strong claims for the finality of our conclusions (see Fig. \ref{ratio}) about
mass composition.  A definitive statement awaits a better understanding of
the differences between different energy spectra above this energy and an
exploration of the sensitivity of the method to different models of particle
interactions.  This technique can be extended to very high energies and,
assuming the interaction model is understood, we would expect to be able to
extract the mass spectrum in the range 10$^{19}$ to 10$^{20}$ eV with data
from the Pierre Auger Observatory.
		
{\bf Acknowledgements:} This work was partly supported by a joint grant from
the British Council and the Spanish Ministry of Education (HB1997-0175), by
Xunta de Galicia (PGIDT00PXI20615PR), by CICYT (AEN99-0589-C02-02) and by
PPARC(GR/L40892).

\end{document}